\documentclass[aps,prl,notitlepage,reprint,superscriptaddress]{revtex4-1}

\usepackage[colorlinks,linkcolor=red,citecolor=blue,urlcolor=red]{hyperref}
\renewcommand{\t}[1]{\mathrm{#1}}

\usepackage{graphicx}
\usepackage{amsmath,amssymb,amsfonts}

\usepackage{amsmath,amssymb,amsfonts} 
\usepackage{bm}	
\renewcommand{\mathbf}{\bm}

\usepackage{dsfont}	
\renewcommand{\mathbb}{\mathds}	

\usepackage{mathrsfs} 
\usepackage{mathtools} 

\usepackage{color}

\newcommand{\fref}[1]{Fig.~\ref{#1}}
\renewcommand{\eqref}[1]{Eq.~\ref{#1}}

\begin{document}
\title{
	Strain engineering for ultra-coherent nanomechanical oscillators}


\author{A. H. Ghadimi}
\thanks{These authors contributed equally to this work}
\affiliation{Institute of Physics (IPHYS), {\'E}cole Polytechnique F{\'e}d{\'e}rale de Lausanne, 1015
	Lausanne, Switzerland}

\author{S. A. Fedorov}
\thanks{These authors contributed equally to this work}
\affiliation{Institute of Physics (IPHYS), {\'E}cole Polytechnique F{\'e}d{\'e}rale de Lausanne, 1015
	Lausanne, Switzerland}

\author{N. J. Engelsen}
\thanks{These authors contributed equally to this work}
\affiliation{Institute of Physics (IPHYS), {\'E}cole Polytechnique F{\'e}d{\'e}rale de Lausanne, 1015
	Lausanne, Switzerland}

\author{M. J. Bereyhi}
\affiliation{Institute of Physics (IPHYS), {\'E}cole Polytechnique F{\'e}d{\'e}rale de Lausanne, 1015
	Lausanne, Switzerland}

\author{R. Schilling}
\affiliation{Institute of Physics (IPHYS), {\'E}cole Polytechnique F{\'e}d{\'e}rale de Lausanne, 1015
	Lausanne, Switzerland}


\author{D. J. Wilson}
\email{dalziel.wilson@epfl.ch}
\affiliation{IBM Research --- Zurich, Sa\"{u}merstrasse 4, 8803 R\"{u}schlikon, Switzerland}

\author{T. J. Kippenberg}
\email{tobias.kippenberg@epfl.ch}
\affiliation{Institute of Physics (IPHYS), {\'E}cole Polytechnique F{\'e}d{\'e}rale de Lausanne, 1015
	Lausanne, Switzerland}

\date{\today}

\begin{abstract}
	
\end{abstract}

\maketitle

{\small
	
{
	\textbf{Elastic strain engineering utilizes stress to realize unusual material properties \cite{li2014elastic}. For instance, strain can be used to enhance the electron mobility of a semiconductor, enabling more efficient solar cells \cite{feng2012strain} and smaller, faster transistors \cite{chidambaram2006fundamentals}.	In the context of nanomechanics, the pursuit of resonators with ultra-high coherence has led to intense study of a complementary strain engineering technique, ``dissipation dilution'', whereby the stiffness of a material is effectively increased without added loss \cite{gonzalez1994brownian,unterreithmeier2010damping,schmid2011damping,tsaturyan2016ultra}.  Dissipation dilution is known to underlie the anomalously high quality factor ($Q$) of $\mathrm{Si_3N_4}$ nanomechanical resonators \cite{southworth2009stress,wilson2009cavity,chakram2014dissipation}, including recently-developed ``soft-clamped'' resonators \cite{tsaturyan2016ultra}; however, the paradigm has to date relied on weak strain produced during material synthesis. By contrast, the use of geometric strain engineering techniques \cite{suess2013analysis,minamisawa2012top} --- capable of producing local stresses near the material yield strength --- remains largely unexplored. Here we show that geometric strain combined with soft-clamping can produce unprecedentedly high $Q$ nanomechanical resonators. Specifically, using a spatially non-uniform phononic crystal pattern, we co-localize the strain and flexural motion of a $\mathrm{Si_3N_4}$ nanobeam, while increasing the former to near the yield strength. This combined strategy produces string-like modes with room-temperature $Q\times$frequency($f$) products approaching $10^{15}$ Hz, an unprecedented value for a mechanical oscillator of any size. The devices we have realized can have force sensitivities of $\text{aN}/\sqrt{\text{Hz}}$, perform hundreds of quantum coherent oscillations at room temperature, and attain $Q>400$ million at radio frequencies. 	
	These results signal a paradigm shift in the control of dissipation in nanomechanical systems, with impact ranging from precision force microscopy \cite{rugar2004single} to tests of quantum gravity \cite{aspelmeyer2012quantum}. 	
	Combining the reported approach with crystalline or 2D materials may lead to further improvement, of as yet unknown limitation.
	}
}

\begin{figure}[t!]
	\includegraphics[width=1\columnwidth]{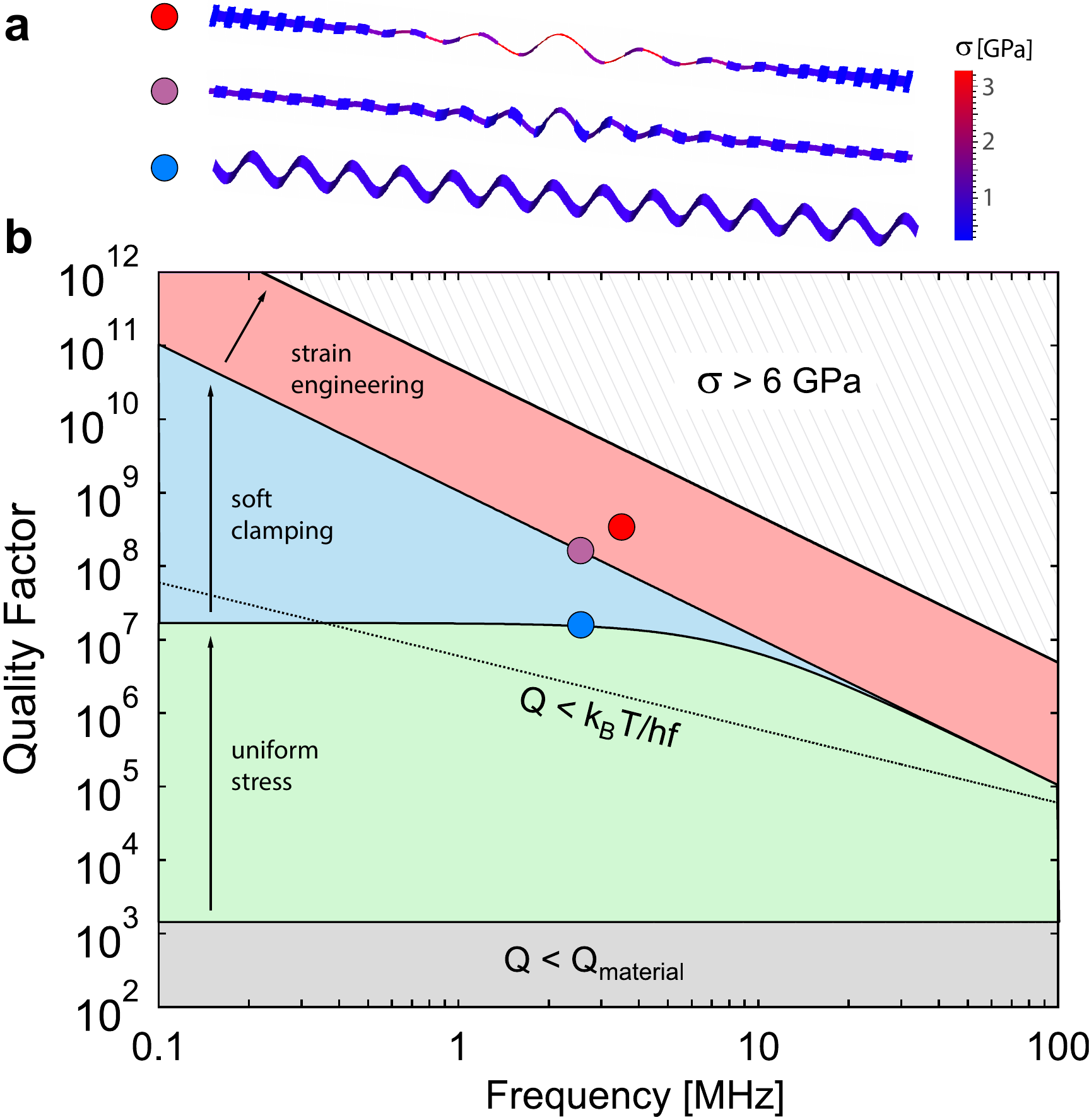}
	\caption{\textbf{Ultra-high-$Q$ nanobeams through dissipation dilution.} (a) Mode shapes representing three strategies to enhance the $Q$ of a nanobeam via dissipation dilution. From bottom to top: uniform stress, soft-clamping, and geometric strain engineering. Color-coding represents axial stress, $\sigma$. (b) $Q$ versus mode frequency $(f)$ accessible for a 20-nm-thick Si$_3$N$_4$ nanobeam, following \eqref{eq:3}. Gray region: $Q(f)$ of an unstressed beam, limited by material loss.  Green region: $Q(f)$ of a 3-mm-long uniform beam with $\sigma < 1$ GPa. Blue region: $Q(f)$ accessible by soft-clamping. Red region: $Q(f)$ accessible by soft-clamping and strain engineering.  Hatched region is forbidden by the material yield strength. Solid circles correspond to measurements described in the main text.}
	\label{figure1}
	\vspace{-3mm}
\end{figure}

Strained nanomechanical resonators have been the subject of intense research over the last decade owing to their anomalously high $Q$ factors \cite{southworth2009stress,wilson2009cavity,chakram2014dissipation}.  The mechanism behind this anomaly, ``dissipation dilution'', is akin to trapping a particle in a conservative potential: strain increases the stiffness of the resonator without adding loss \cite{gonzalez1994brownian,unterreithmeier2010damping}.  Unlike most bulk mechanical properties, dissipation dilution can scale inversely with device dimensions, implying that smaller mass ($m$) devices can have higher $Q$.  This unusual scaling is highly favorable for sensing applications, in which $m^{-1}$ dictates the motional response and $Q^{-1}$ the scale of environmental noise \cite{saulson1990thermal}.  Adding to this intrigue, it has been shown that nanoresonators made of high stress material ($\sigma\gtrsim 1$ GPa) can achieve $Q\times f$ products exceeding $6\times 10^{12}$ Hz, satisfying the basic requirement for quantum coherence at room temperature (viz., in this case the thermal decoherence time, $hQ/k_\t{B}T$, exceeds one mechanical period) \cite{wilson2009cavity}.  Operated at moderate cryogenic temperatures, such devices have enabled several landmark demonstrations of ``macroscopic'' quantum effects (notably in the field of cavity optomechanics) \cite{purdy_observation_2013,purdy_strong_2013,wilson2015measurement}.  

 \begin{figure*}[t!]
	\includegraphics[width=2\columnwidth]{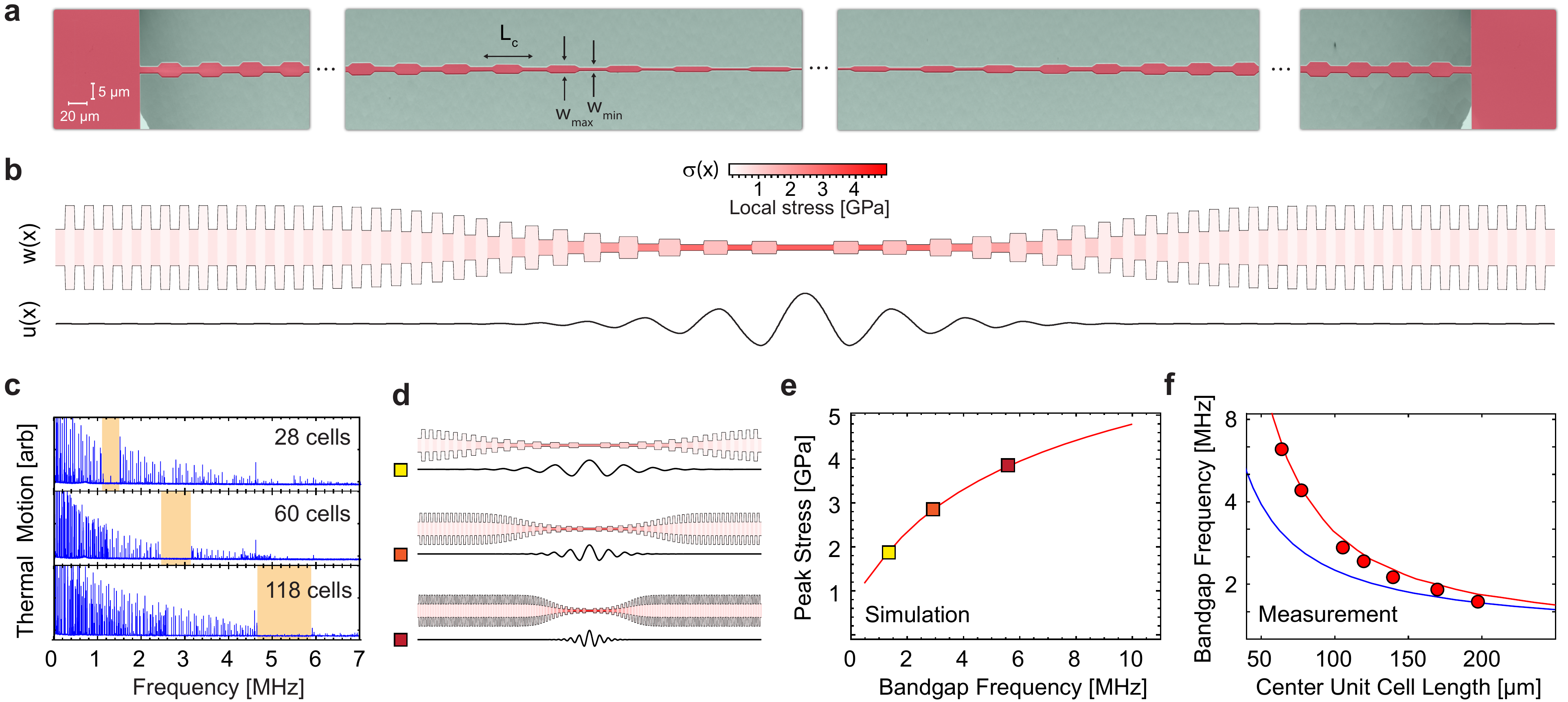}
	\caption{\color{black}\textbf{Strain-engineered 1D phononic crystals.} (a) Scanning electron micrograph (SEM) of a typical device, vertically scaled for perspective. (b) Width/stress profile and defect mode shape of a device with 60 unit cells. (c) Thermal displacement spectrum of beams with profiles shown in (d).  Bandgaps are highlighted in orange. (e) Simulation of peak stress versus bandgap frequency $f_\t{bg}$ for devices shown in (d).  (f) Measurements of $f_\t{bg}$ versus length of the central unit cell (parameterizing the taper length). Red and blue lines are models with and without accounting for stress localization, respectively (see text). }
	\label{figure2}
	\vspace{-3mm}
\end{figure*}

Remarkably, the potential of strained nanomechanical resonators seems largely untapped. Despite impressive advances in the use of mode-shape-engineering to enhance dissipation dilution, recently resulting in radio frequency resonators with room temperature $Q\times f\gtrsim10^{14}$ Hz 
\cite{tsaturyan2016ultra,norte2015mechanical,chakram2014dissipation}, the vast majority of studies have focused on devices fabricated from amorphous Si$_3$N$_4$ under weak deposition stress. Fruitful lessons might be learned from the field of elastic strain engineering \cite{li2014elastic}, where extreme geometric stresses, often approaching the material yield strength \cite{suess2013analysis,minamisawa2012top}, are used to control the electronic properties of crystalline films and 2D materials.

Inspired by this connection, here we explore a technique to achieve ultra-coherent nanomechanical resonators by combining mode shape engineering and geometric strain. 
Our approach, visualized in \fref{figure1}, is conceptually simple and entirely material independent: By weakly corrugating a pre-stressed nanobeam, we create a bandgap for localizing its flexural modes around a central defect. By tapering the beam, we co-localize these modes with a region of enhanced stress. Reduced motion near the supports (``soft-clamping'' \cite{tsaturyan2016ultra}) results in 
higher dissipation dilution, while enhanced stress increases both dilution and mode frequency. 
Leveraging a multi-step release process, we implement our approach on extremely high aspect ratio tapered beams (as long as 7 mm and as thin as 20 nm) made of pre-stressed (1 GPa) Si$_3$N$_4$, and achieve geometric stresses as high as 4 GPa. Megahertz-frequency modes of these beams are found to exhibit room temperature $Q$ factors as high as $4\times10^{8}$ and $Q\times f$ products as high as $8\times 10^{14}$ Hz.  Owing to their picogram-scale effective masses, these modes are exquisite force sensors, limited by a thermal noise floor of less than $10\,\t{aN}/\sqrt{\t{Hz}}$ \cite{saulson1990thermal}.  Owing to their large zero-point motion and low thermal decoherence rates ($k_B T/hQ\sim 10$ kHz), they are also promising as macroscopic quantum systems.  They perform hundreds of quantum-coherent oscillations at room temperature, and should exhibit sub-kHz decoherence rates, on par with trapped ions \cite{turchette2000heating}, at 4 K.

To illustrate the core concepts behind our approach, we first consider a model for dissipation dilution of a \emph{non-uniform} beam of length $L$, thickness $h$ and variable width $w(x)$. Following an anelastic approach successfully applied to uniform nanobeams \cite{unterreithmeier2010damping,schmid2011damping} and nanomembranes \cite{yu2012control}, we partition the potential energy of the beam $U$ into two components: a dissipative component due to bending, $U_E=\tfrac{1}{2}E_0 \int_0^{L} I(x) [u''(x)]^2dx$, and a conservative component due to elongation, $U_\sigma=\tfrac{1}{2}T\int_0^{L} [u'(x)]^2dx$, where $u(x)$ is the vibrational modeshape and $I(x)= \tfrac{1}{12}w(x)h^3$, $T=h w(x)\sigma(x)$ and $\sigma(x)$ are the geometric moment of inertia, tension, and axial stress of the beam, respectively.  The $Q$ enhancement due to stress (the dissipation ``dilution factor") is given by the participation ratio of the lossy potential \cite{gonzalez1994brownian,schmid2011damping,unterreithmeier2010damping}:
\begin{equation}\label{eq:1}
\frac{Q}{Q_0}=1+\frac{U_\sigma}{U_E}\approx\frac{12}{E_0 h^2}\cdot\frac{\int_0^{L} [u'(x)]^2dx}{\int_0^{L} \sigma^{-1}(x) [u''(x)]^2dx},
\end{equation}
where $Q_0$ is the intrinsic (undiluted) quality factor.

For the familiar case of a uniform beam with string-like modeshape $u(x)\propto \sin(\pi n x/L)$, \eqref{eq:1} implies that
\begin{equation}\label{eq:2}
\frac{Q}{Q_0}\approx\frac{3}{\pi^2}\frac{\sigma^2}{E_0 h^2\rho f^2},
\end{equation}
where $\rho$ is the material density and $f=\tfrac{n}{L}\sqrt{\sigma/\rho}$ is the mode frequency. Remarkably, nearly all stressed nanomechanical resonators studied to date have operated far below this limit.  
The main reason for this discrepancy is clamping loss, e.g., for a doubly-clamped beam, boundary conditions $u'(x_0)=u(x_0)=0$ require that the vibrational modeshape exhibit extra curvature ($u''(x)$) near the supports ($x_0=0,L$), resulting in a reduced dilution factor of the form


\begin{equation}\label{eq:3}
\frac{Q}{Q_0}\approx (\hspace{-4pt}\underbrace{2\lambda}_\t{supports}+\underbrace{\pi^2n^2\lambda^2}_\t{antinodes})^{-1},
\end{equation}
where $\lambda = \tfrac{h}{L}\sqrt{E_0/12\sigma}$ \cite{villanueva2014evidence}.



 \begin{figure*}[t!]
 	\includegraphics[width=2.05\columnwidth]{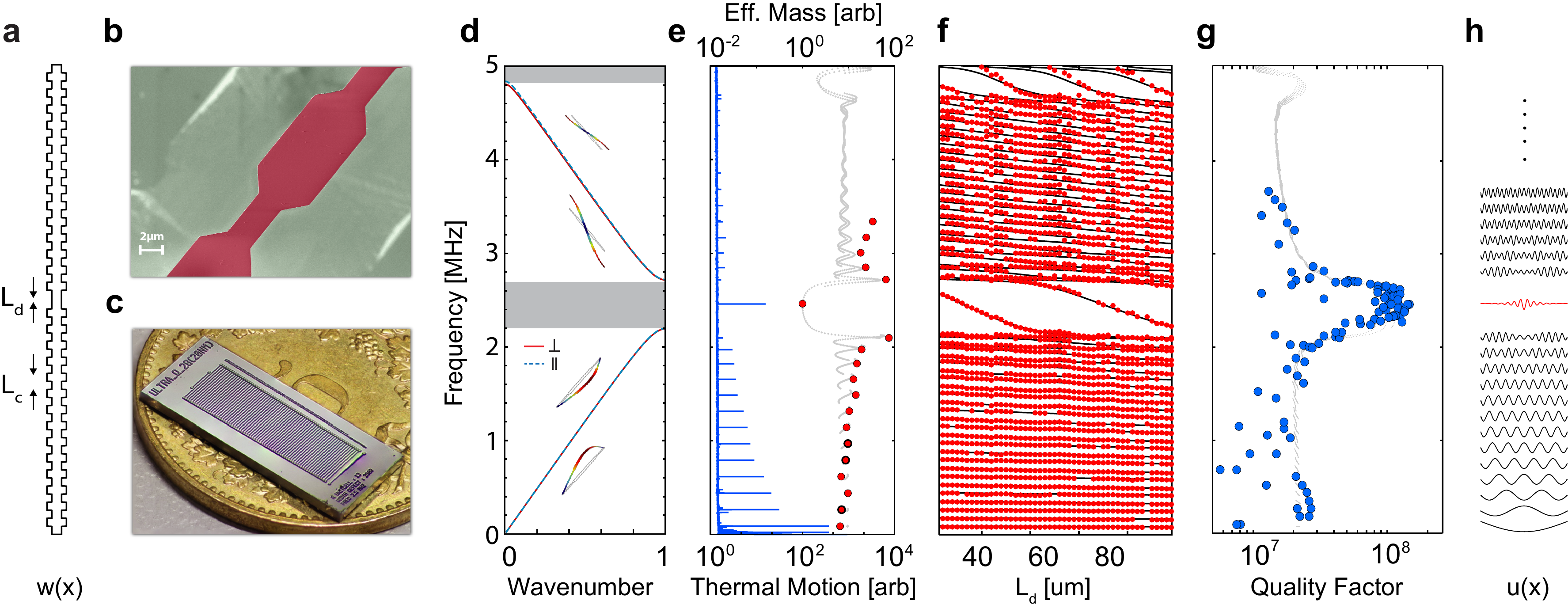}
 	\caption{\textbf{``Soft-clamped" 1D nanomechanical resonators.} (a)  Schematic of a phononic crystal nanobeam with central defect. See main text for details. (b) SEM image of a unit cell.  (c) Optical image of a sample chip with 76 beams, each with a different defect length. (d) Band diagram showing in($\parallel$)- and out($\perp$)-of-plane  normal modes of a unit cell. (e) Thermal displacement spectrum (blue) of a single beam, overlaid with effective mass coefficients (red circles) inferred from the area beneath noise peaks.  Gray is a model based on modeshapes in (h) (see Methods). (f) Frequency spectra of multiple beams with different defect length. Black lines are a solution to the Euler-Bernoulli equation.  (g) Compilation of $Q$ measurements for a subset of the modes in (f), overlaid with a model based on \eqref{eq:1}. (h) Modeshapes obtained from the Euler-Bernoulli equation.}
 	\label{figure3}
 	\vspace{-3mm}
 \end{figure*}

The uniform beam model (\eqref{eq:3}) gives several rules of thumb for maximizing the $Q$ or $Q\times f$ of a stressed nanomechanical resonator, namely, $Q$ is typically highest for the fundamental mode ($n=1$) and can be increased by increasing aspect ratio ($L/h$) or stress.  By contrast, $Q\times
 f$ is typically larger for high order modes.  Both strategies have been explored for a wide variety of beam and membrane-like geometries \cite{villanueva2014evidence,ghadimi2017radiation}.  
 A third approach, recently demonstrated by Tsaturyan \emph{et. al.} \cite{tsaturyan2016ultra} with a membrane, is to use periodic micropatterning (a phononic crystal (PnC)) to localize the modeshape $u(x)$ away from the supports. 
 By this ``soft-clamping" approach, the leading term in \eqref{eq:3} can be suppressed, giving access to the performance of an ideal clamp-free resonator (\eqref{eq:2}).

Complementary to soft-clamping, our approach consists of co-localizing modeshape   $u(x)$     with a region of geometrically-enhanced stress $\sigma(x)$, making use of the tension balance relation, $\sigma(x)=T/(w(x)h)$. 
Geometric stress has been exploited in the past to enhance the $Q\times f$ of a nanomechanical resonator \cite{zhang_nanoscale_2016}; however, performance was in this case limited by rigid clamping.
Combining geometric stress and soft-clamping can lead to dramatically enhanced performance: For example, \eqref{eq:2} suggests that the $Q$ (for a fixed $f$) of a typical 1 GPa pre-stressed Si$_3$N$_4$ nanobeam can be enhanced by a factor of 50 before the stress in the thinnest part of the beam reaches the yield strength of Si$_3$N$_4$ ($\sigma_\t{yield}\approx6$ GPa).  This material limit, described by \eqref{eq:2} with $\sigma=\sigma_\t{yield}$ and illustrated by the hatched region in \fref{figure1}, can be shown to apply to an arbitrary beam profile $w(x)$ (see Methods). 
In gaining access to it, the main caveat of our approach is the small spatial extent of the enhanced stress, which implies that high order flexural modes must be used to achieve sufficient co-localization.


To explore the above concepts, devices were patterned on 20-nm-thick films of high-stress Si$_3$N$_4$ ($E_0\approx 250$ GPa, $\sigma_0 \approx 1.1$ GPa) grown by low pressure chemical vapor deposition on a Si wafer.
A multi-step release process (see Methods) was employed to suspend beams as long as 7 mm, enabling aspect ratios as high as $3.5\times10^5$ and dilution factors in excess of $(2\lambda)^{-1}\approx 3\times 10^4$. 
To realize PnCs, beams were corrugated with a simple step-like unit cell of length $L_\t{c}$, minor width $w_\t{min}$, and major width $w_\t{max}\approx 2w_\t{min}$ (see \fref{figure2}a). A uniform defect of length $L_\t{d}$ was patterned at the center of each beam in order to define the position of localized modes. To co-localize stress with these modes, we explore a strategy whereby the width of successive unit cells is adiabatically tapered toward the defect according to a Gaussian envelope function (see Methods). 

It is worth noting that localized modes of PnC nanobeams (``1D phononic crystals") have already been widely studied, as their ultra-low-mass and sparse mode spectrum makes them highly promising for sensing applications.  In contrast to 2D (membrane-like) resonators \cite{tsaturyan2016ultra}, however, ultra-high-$Q$ in 1D PnCs has not been reported to date, due to a focus on unstrained materials \cite{chan2012optimized}, and/or highly-confined (high curvature) modes \cite{ghadimi2017radiation} limited by radiation loss.  With this discrepancy in mind, we first embarked on a study of \emph{uniform} (untapered) PnC nanobeams, focusing on weakly-localized modes of our high-aspect-ratio devices.  

\begin{figure*}[t!]		\includegraphics[width=1.85\columnwidth]{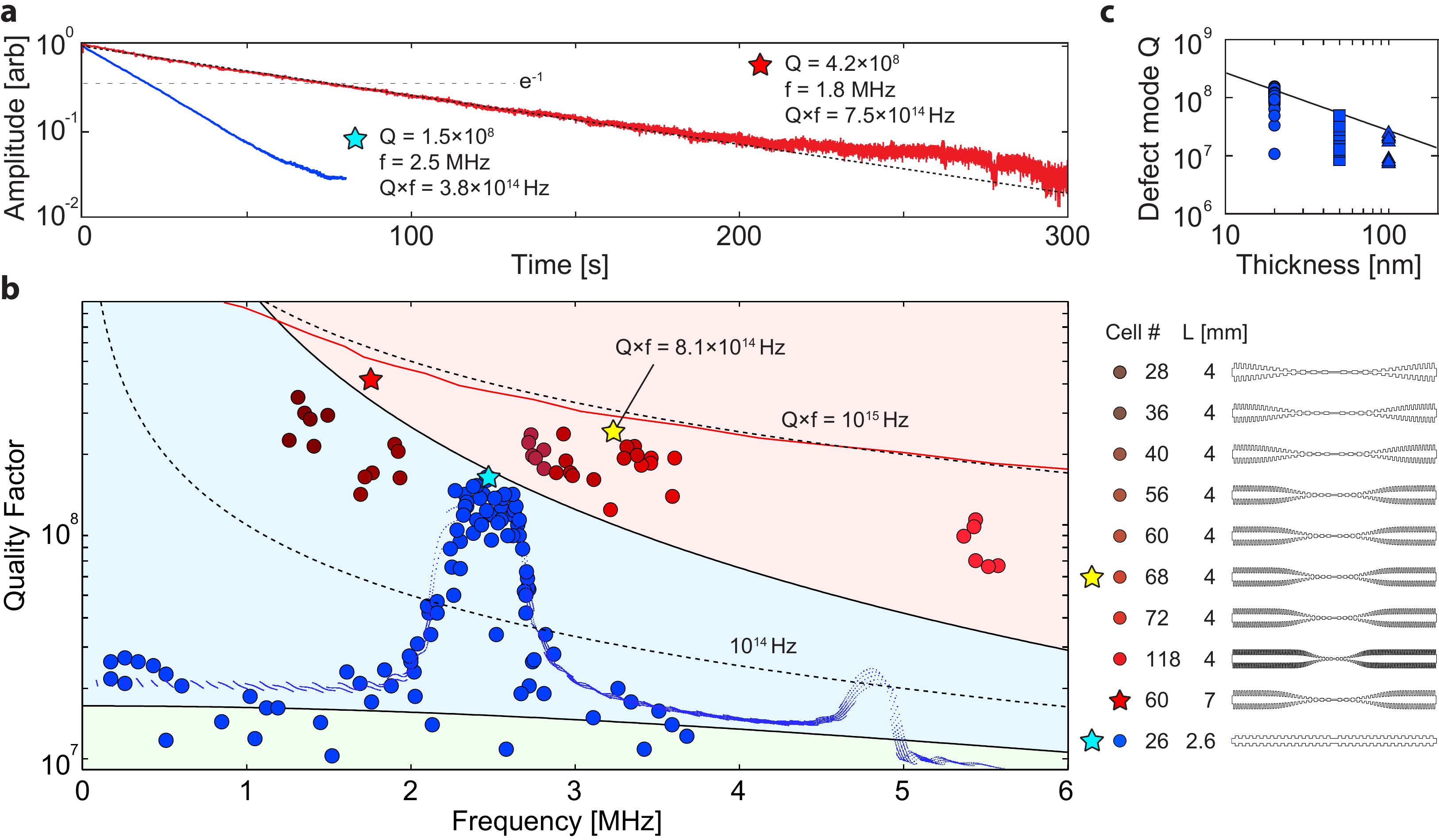}
	\caption{\color{black}\textbf{Enhancing the quality factor a soft-clamped nanobeam by strain-engineering.} (a) Ringdown of a 7-mm-long, 20-nm-thick tapered PnC nanobeam excited in its 1.76 MHz defect mode (red).  Dotted line is an exponential fit to a lifetime of 76 seconds. The inferred $Q$ of $4.2\times 10^8$ is indicated by a star in (b). Also shown is a ringdown of a uniform PnC nanobeam (cyan star in (b)).  (b) $Q$ versus mode frequency of PnC nanobeams with different geometries, summarized at right.  Blue points correspond to modes of the uniform PnC nanobeam described in \fref{figure3}.  Red points correspond to defect modes of tapered beams.  Yellow star indicates the highest $Q\times f$ product measured. Each color group includes the highest five $Q$s recorded for multiple beams. Red lines and blue dots are numerical models based on \eqref{eq:1}. (c) Compilation of defect mode $Q$ for uniform PnC nanobeams (\fref{figure3}) of different thickness.  Overlaid is a model with $Q_0 = 6900\cdot h/(100\,\t{nm})$, consistent with surface loss.}
	\label{figure4}
	\vspace{-3mm}
\end{figure*}

An experiment demonstrating soft-clamped 1D nanomechanical resonators is presented in \fref{figure3}.  2.6-mm-long devices with unit cells of length $L_c = 100\,\mu$m and width $w_\t{min(max)}=0.5(1)\,\mu$m were studied.  To characterize these devices, thermal noise and ringdown measurements were carried out \emph{in vacuo} ($<10^{-6}$ mbar) using a custom lensed fiber interferometer (see Methods). 
As an attractive consequence of their simple geometry, mode frequencies (inferred from thermal noise spectra, \fref{figure3}e) were found to agree strikingly well with a numerical solution to the 1D Euler-Bernoulli (E-B) equation (
see Methods).  Also striking is the sparse mode spectrum inside  the bandgap, visualized in \fref{figure3}f by compiling spectra of beams with different defect lengths.  A single defect mode appears to move in and out of the bandgap in \fref{figure3}f as the defect length is varied.  This mode is expected to be localized and to there have a reduced effective mass, $m$. 
Comparing the area under thermal noise peaks 
and estimating the physical beam mass to be $m_0=100$ pg, 
we infer that indeed $m\approx5\,\t{pg}\ll m_0$ (see Methods).  This value is in good agreement with the mode profile obtained from the E-B equation (\fref{figure3}g), and is roughly 2 orders of magnitude smaller than that of an equivalent 2D localized mode.

In accordance with \eqref{eq:3}, we also observe a dramatic increase in the $Q$ of localized modes.  To visualize this enhancement, we compiled measurements of $Q$ versus mode frequency for $40$ beams of different defect length (\fref{figure3}g). Outside the bandgap, we find that $Q(f)$ is consistent with that of a uniform beam, asymptoting at low mode order ($n\lesssim20$) to $Q\approx 2\times 10^7$, implying $Q_0 \approx 2\lambda Q \approx 1500$. Inside the bandgap $(n\approx 26)$, $Q$ approaches that of an idealized clamp-free beam ($Q\approx Q_0/(\pi  n \lambda)^2\approx 10^8$). The transition between these two regimes is in good agreement with a full model (gray dots) obtained using \eqref{eq:1}.
In \fref{figure4}a we highlight the $19$ second amplitude ringdown of a 2.46 MHz defect mode, corresponding to $Q= 1.5\times10^8$ and $Q\times f = 3.7\times 10^{14}$ Hz.  
Combined with its low effective mass, the room temperature force sensitivity of this mode is an estimated to be $\sqrt{8\pi k_B Tm f/Q}\approx 3\,\t{aN}/\sqrt{\t{Hz}}$, on par with a typical AFM cantilever operating at 100 times lower frequency and temperature \cite{poggio_feedback_2007}.


Having established near-ideal soft-clamping of uniform nanobeams, we next studied the performance of strain-engineered (tapered) nanobeams. Towards this end, a set of 4-mm-long tapered PnC nanobeams was fabricated with the length of the taper varied in order to tune the stress at the center of the beam $\sigma(x_c)$ from 2 to 4 GPa (\fref{figure2}d-e).  We note that for our tapering strategy the width of beam center $w(x_c)$ is fixed, so that the stress is tuned by changing the equilibrium tension $T$ (see Methods). Moreover, for each taper length we engineer the soft-clamped mode to be well localized inside the thin taper region by tuning the density of unit cells.
As shown in \fref{figure2}e, measurements of bandgap frequency $f_\t{bg}$ versus length of the central unit cell length $L_{c,0}$ (parameterizing the taper length) corroborate enhanced stress through correspondence with the theoretical scaling $f_\t{bg}\propto\sqrt{\sigma(x_c)}/L_{c,0}$.

The central results of this report are shown \fref{figure4}, in which the $Q$ factors of uniform and tapered PnC nanobeams are compared.  Blue points correspond to the measurements in \fref{figure3}h.  Red points are compiled for the localized modes of tapered beams with various peak stresses, corresponding to a bandgap frequency varied from $f_\t{bg}=1-6$ MHz.  According to a full model (red line), $Q(f_\t{bg})$ should in principle trace out a contour of constant $Q\times f\approx 10^{15}$ Hz, exceeding the clamp-free limit of a uniform beam ($Q\times f\propto 1/f$) for sufficiently high frequency.  We observe this behavior with an unexplained $\sim 30\%$ reduction, with $Q$ factors exceeding the clamp-free model by a factor of up to three and reaching absolute values high as $3\times 10^8$. Though theoretically this $Q$ should be accessible by soft-clamping alone at lower frequency, our strain-engineering strategy gives access to higher $Q\times f$, reaching a value as high as $8.1\times 10^{14}$ Hz for the $3.2$ MHz mode of a 4-mm-long device (yellow star) and $Q$ as high as $4.2\times10^8$ for a the 1.8 MHz mode of a 7-mm-long device (red star).  To our knowledge this is the highest $Q$ on record for a nano- or micromechanical resonator at room temperature, and the highest $Q\times f$ at room temperature for a mechanical resonator of any size.

Several aspects of these results are worth qualifying.  First, the dilution factors we have achieved are still an order of magnitude below the limit set by the breaking stress of Si$_3$N$_4$.  Our results may thus benefit from more aggressive strain engineering. (For example, 30-nm-wide Si microbridges have been fabricated with geometric stresses as high as 7.6 GPa \cite{minamisawa2012top}.) 
We also emphasize that higher aspect ratio devices offer a direct route to higher $Q$.  The aspect ratios of our longest beams ($L/h=3.5\times10^5$ for a $L=7$ mm device) appear to be anomalously high for a suspended thin film, including 2D materials; however, a recent report has demonstrated high-stress Si$_3$N$_4$ membranes with cm-scale dimensions \cite{moura2017centimeter}, hinting at a trend towards more extreme devices.  Finally, we note that the source of internal loss in our devices is unknown, but likely due to surface imperfections \cite{villanueva2014evidence}.  To test this hypothesis, defect-$Q$s for beams with thicknesses $h=$ 20, 50, and 100 nm were compiled (\fref{figure4}c). The inferred thickness dependence of the internal $Q$, $Q_0\approx 6900\cdot h/100\,\t{nm}$, is indeed a signature of surface loss, and agrees well in absolute terms with a recent meta-study on Si$_3$N$_4$ nanomechanical resonators \cite{villanueva2014evidence}.  Interestingly, the $Q\propto Q_0/h^{2}$ scaling of soft-clamped resonators \cite{tsaturyan2016ultra} preserves the advantage of thinner devices even in the presence of this ubiquitous loss mechanism.

Looking forward, a number of directions seem promising for realizing yet higher $Q$. One route is to fabricate mechanical resonators from strained 2D materials. Extreme aspect ratios as high as $3\times 10^5$ have been demonstrated for suspended graphene sheets \cite{barton_high_2011}, matching the highest values realized in the present work.  These materials can moreover have yield stresses well in excess of 10 GPa \cite{lee2008measurement}. 
Another route is to reduce intrinsic loss, for instance by improved surface conditioning \cite{villanueva2014evidence,tao2014single} or by employing crystalline thin films at cryogenic temperatures \cite{cole2013tenfold,tao2014single}. The latter approach is intriguing because pre-stresses of $\sigma_0\sim$ GPa are in fact readily accessible by lattice mismatch in epitaxial growth (for critical thicknesses of $h\sim10$ nm). 
To give an example: engineering the reported devices out of a 1$\%$-strained InGaP film ($\sigma_0\approx 1$ GPa, $E_0\approx 100$ GPa) \cite{cole2014tensile} and operating at 4 K should in principle enable $Q_0\sim 10^5$, $Q\sim10^{11}$ and $Q\times f\sim10^{17}$ Hz, corresponding to zeptonewton force sensitivities and an astoundingly low thermal decoherence rate of $k_B T/hQ\sim 1$ Hz.  These values spark the imagination, inviting speculation that mechanical oscillators might one day serve as a platform for solid state quantum sensing \cite{degen2017quantum}, in combination with or in lieu of artificial atoms.

\vspace{-2mm}
\section{Acknowledgements}
The authors thank Hendrik Sch\"{u}tz and Ehsan Mansouri for valuable contributions during the initial phase of the experiment.  This work was supported by the EU Horizon 2020 Research and Innovation Program under grant agreement No. 732894 (FET Proactive HOT) and the SNF Cavity Quantum Optomechanics project (grant no. 163387). M.J.B. is supported by MSCA ETN-OMT (grant no. 722923). T.J.K  acknowledges support from ERC AdG (QuREM, grant no. 320966). All samples were fabricated at the Center for MicroNanoTechnology (CMi) at EPFL.
\color{black}
\vspace{-2mm}



%

\clearpage


\setcounter{equation}{0}
\setcounter{figure}{0}
\setcounter{table}{0}
\onecolumngrid
\vspace{1cm}
\begin{center}
	
	\Large \textbf{Methods for \\``Strain engineering for ultra-coherent nanomechanical oscillators"}
\end{center}
\vspace{0.9cm}
\twocolumngrid

\small{
	\section{Calculation of vibrational spectra}
	For the modeling of the vibrations of nanobeam resonators we use the 1D Euler-Bernoulli equation,
	\begin{equation}\label{eq:EulerBernoulli}
	\frac{d^2}{dx^2}\left(EI(x)\frac{d^2 u}{dx^2}\right)-T\frac{d^2 u}{dx^2}-\rho_x(x)\omega^2 u=0,
	\end{equation}
	which is adequate for the description of our structures due to their extreme aspect ratios $L\gg h,w$. Here $E$ is the Young's modulus of the material, $u(x)$ and $\omega$ are the vibrational mode shape and frequency respectively and $\rho_x=\rho h w(x)$ is the linear beam density. The equilibrium tension $T$ is related to the deposition stress $\sigma_\t{dep}$ as
	\begin{equation}
	T=\sigma_\t{dep}(1-\nu)\left(\frac{1}{L}\int_{0}^{L}\frac{1}{w(x)}dx\right)^{-1},
	\end{equation}
	where $\nu$ is the Poisson ratio. Variation of the axial stress along the beam is given by $\sigma(x)=T/(h\,w(x))$. In the calculations we use the following numerical values for the material parameters: $E=250$ GPa, $\nu=0.23$ and $\sigma_\t{dep}=1.1$ GPa.\par
	We solve \eqref{eq:EulerBernoulli} with clamped boundary conditions ($u(0)=u(L)=0,\,u'(0)=u'(L)=0$) to find the frequency spectra of the utilized beam structures. The phononic band diagram in Fig. 3d is generated by solving the equation for one unit cell with periodic boundary conditions ($u(x+L_\t{cell})=u(x)e^{i\theta_F}$).\par
	The effective masses plotted in Fig. 3E are defined with respect to a point probe at the beam center ($x=x_c$):
	\begin{equation}\label{eq:mEff}
	m_\t{eff}=\rho h\frac{1}{u(x_c)^2}\int_{0}^{L}w(x)u(x)^2dx.
	\end{equation}
	For the fundamental flexural mode of a non-tapered phononic crystal beam according to \eqref{eq:mEff} $m_\t{eff}=m_\t{phys}/2$, where $m_\t{phys}=\rho h\langle w\rangle L$ is the physical mass of the beam. Experimentally, the effective masses shown in Fig. 3e were inferred relative to the fundamental mode $m_\t{eff}$ from the Brownian motion spectrum, where the areas under individual vibrational peaks are proportional to the mean-squared position fluctuations, which are proportional to $ k_B T/(m_\t{eff} f^2)$.
	
	\section{Upper bound for loss dilution of a non-uniform beam}
	This section analytically derives the absolute maximum loss dilution achievable with any transverse profile of the beam and shows that it coincides with Eq. 2 in the main text assuming $\sigma=\sigma_\t{yield}$. We first assume that eigenmode shapes and frequencies are well-approximated by solutions to the wave equation, which is always valid in the regime of strong loss dilution:
	\begin{equation}\label{eq:waveEq}
	u_n''(x)+\frac{\rho\, h\, w(x)}{T}\omega_n^2 u_n(x)=0.
	\end{equation}
	To upper-bound the bending energy of mode $n$,
	\begin{equation}
	U_\t{bend}=\frac{1}{2}E_0 \int_0^{L} I(x) [u_n''(x)]^2dx,
	\end{equation}
	we use $\sigma(x)=T/(h\,w(x))$ to rewrite \eqref{eq:waveEq}, giving
	\begin{subequations}\begin{align}
		I(x) [u''(x)]^2&=\frac{h^3 w(x)}{12}\left(-\frac{\rho_0}{\sigma(x)}\omega_n^2 u(x)\right)u''(x)\\
		&\ge \frac{h^2\omega_n^2\rho}{12\,\sigma_\t{peak}^2}T\left(-u(x)\right)u''(x),
		\end{align}\end{subequations}
	where $\sigma_\t{peak}$ is the peak axial stress.
	
	\begin{figure}[t!]
		\includegraphics[width=1\columnwidth]{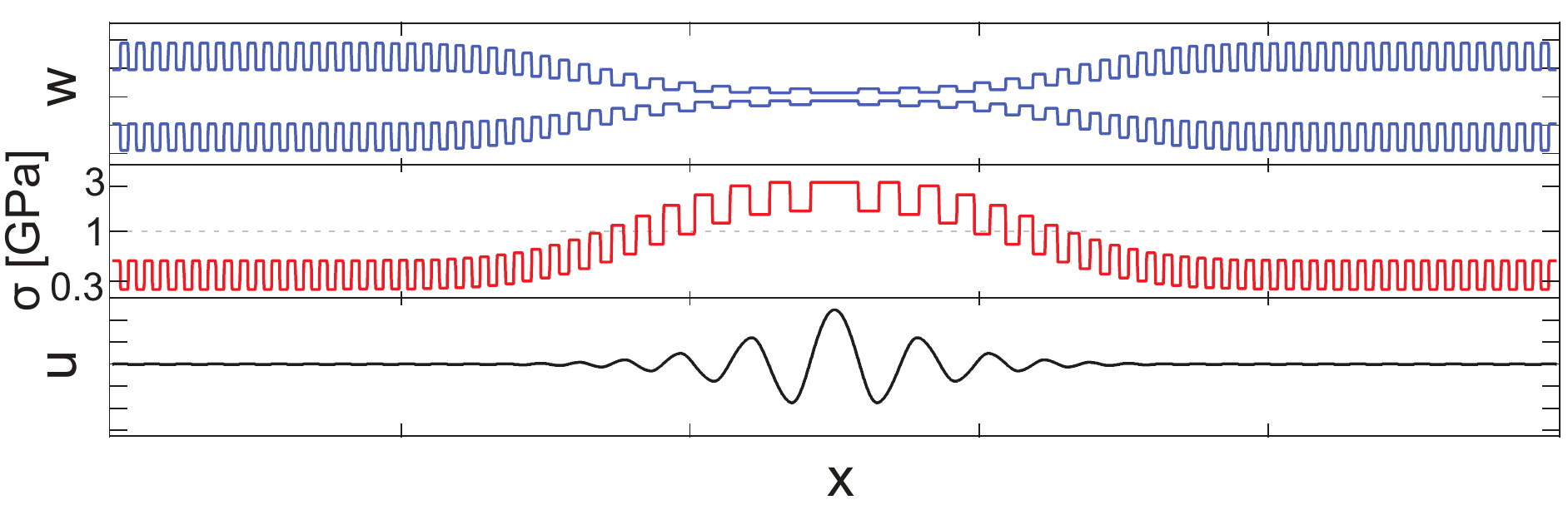}
		\label{figureBeamDesign}
		\caption{\textbf{Design of tapered 1D PnC.} Details of the device shown in Fig. 2a of the main text: Width profile $w(x)$, axial stress $\sigma(x)$, and out-of-plane displacement $u(x)$ of the first order localized mode, plotted versus axial coordinate along the beam, $x$. A deposition stress of $\sigma_\t{dep} = 1.1$ GPa is assumed.}
	\end{figure}
	
	Integration by parts together with the boundary conditions $u(0)=u(L)=0$ gives
	\begin{equation}
	\int_0^{L} I(x) [u''(x)]^2dx\ge\frac{h^2\rho\,\omega_n^2}{12\,\sigma_\t{peak}^2}T\int_0^{L} [u'(x)]^2dx
	\end{equation}
	and therefore
	\begin{equation}
	U_\t{bend}\ge \frac{E\,h^2\rho\,\omega_n^2}{12\sigma_\t{peak}^2}U_\t{elong}.
	\end{equation}
	
	Requiring that $\sigma_\t{peak}$ not exceed the yield stress $\sigma_\t{yield}$ gives the upper bound on loss dilution highlighted in Fig. 1:
	\begin{equation}\label{eq:lossDilLim}
	\frac{U_\t{elong}}{U_\t{bend}}\le \frac{3}{\pi^2} \frac{\sigma_\t{peak}^2}{E\,h^2\rho f_n^2}\le \frac{3}{\pi^2} \frac{\sigma_\t{yield}^2}{E\,h^2\rho f_n^2}.
	\end{equation}
	We emphasize that in deriving \eqref{eq:lossDilLim} from   \eqref{eq:waveEq}, we neglect mode curvature at the clamping points. This is justified because additional curvature would only increase $U_\t{bend}$.
	
		\section{Design of tapered beams}
	Tapering of the beam width $w(x)$ is applied cell-wise according to Gaussian envelope function
		\begin{equation}
		w_\t{max}(i)=2.3\cdot w_\t{min}(i)\propto 1-(1-\alpha_w)\exp(-i^2/i_0^2),
	\end{equation}
	where $i=0,1, ...$ is the unit cell number counting from the central defect, $\alpha_w=0.15-0.2$, and $i_0=8-10$. $\alpha_w$ and $i_0$ were optimized by a random-search algorithm to maximize $Q/Q_0$. Importantly, we also taper the unit cell length as $L_c(i)\propto 1/\sqrt{w_\t{max}(i)}$.  This has the effect of matching the bandgap frequency of each cell, ensuring strong co-localization of stress and defect motion.
	
	\section{Measurement Apparatus}
	Nanobeams were characterized using a custom balanced homodyne interferometer whose signal arm is terminated with a lensed fiber (see Fig. S1 for details).  
	To align a nanobeam to the interferometer, the sample chip is maneuvered in the focal plane of the lens using a 3-axis nanopositioning stage.  Stage, fiber lens, and sample chip are housed together in a high vacuum chamber in order to minimize gas damping.
	
	For the measurements reported, a signal (reference) arm power of $0.1 - 10$ (1) mW was used. The reference arm length was servoed with a piezo-actuated mirror in order to stabilize the interference fringe at  its inflection point. Despite alignment inefficiency (typically $0.1\%$ of the power incident on the beam was retro-reflected into the fiber), thermal motion of the nanobeam could be observed in the power spectrum of the photocurrent at frequencies as high as 10 MHz (Fig. 2).
	
	\begin{figure}[t!]
		\includegraphics[width=1\columnwidth]{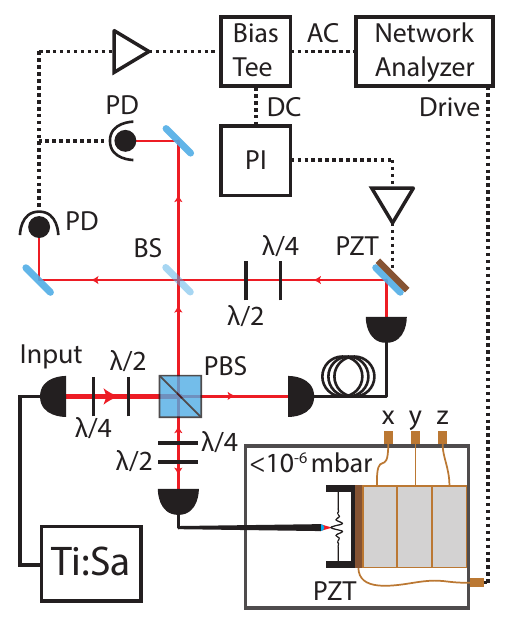}
		\caption{\textbf{Schematic of measurement setup.} $\lambda/2$: Half-wave plate. $\lambda/4$: Quarter-wave plate. PBS: Polarizing beamsplitter. BS: 50/50 beamsplitter. PZT: Piezoelectric transducer. PI: Proportional-integral loop filter. PD: Photodiode.   }
		\label{figure1}
		\vspace{-3mm}
	\end{figure}

	Ringdown measurements were initiated by resonantly exciting the nanobeam (at frequency $f$) using a thin piezo-electric plate attached to the sample holder. After turning off the drive, the slowly-varying amplitude of the beam was monitored by demodulating the photocurrent at $f$ with a bandwidth $\t{BW} \gg f/Q$. Ringdown times were found to be independent of signal power in the range $0.1-10$ mW, suggesting that photothermal damping was negligible. The smallest damping rates reported in Fig. 3 ($f/Q\approx 3$ mHz) were measured at a gas pressure of less than $5\times 10^{-8}$ mbar.  We estimate the contribution of gas damping for this measurement (made on a 20-nm-thick beam) to be less than 10\%. 

	\begin{figure}[b!]
		\includegraphics[width=1\columnwidth]{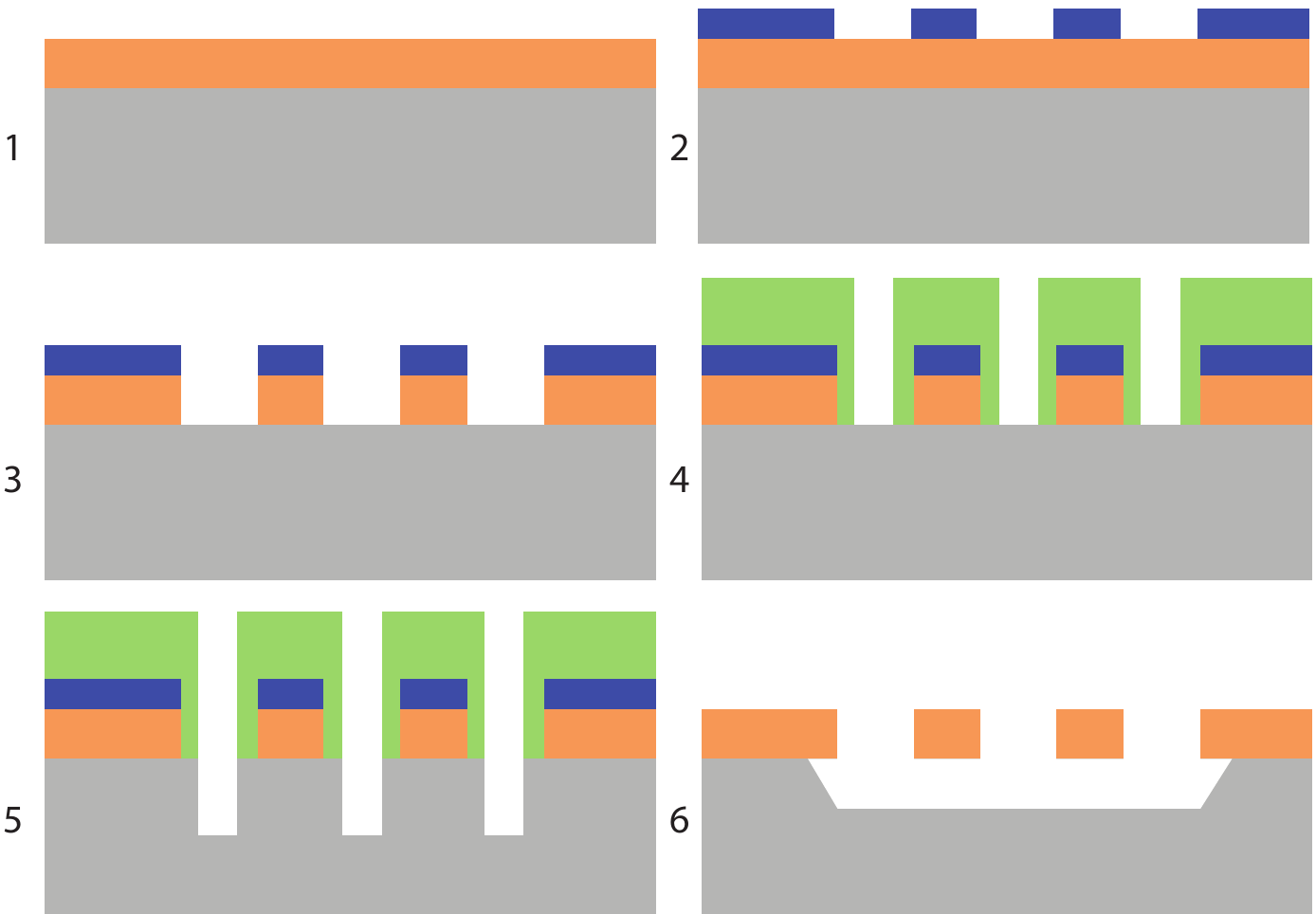}
		\caption{\textbf{Fabrication process flow.} Color-coding: Gray = $\mathrm{Si}$, red = $\mathrm{Si_3N_4}$, blue = first HSQ layer for the main mask, green = second HSQ layer for upscaled mask.}
		\label{figure2}
		\vspace{-3mm}
	\end{figure}

	\section{Fabrication}
	
	The fabrication process (see \fref{figure2} for details) starts with low pressure chemical vapor deposition (LPCVD) of stoichiometric $\mathrm{Si_3N_4}$ on a Si substrate (1). The main processing steps are patterning beams atop the $\mathrm{Si_3N_4}$ layer by electron beam (e-beam) lithography with a hydrogen silsesquioxane (HSQ) resist (2), transferring the patterns to the $\mathrm{Si_3N_4}$ layer by reactive ion etching (RIE) using Florine chemistry (3) and releasing beams from the underlying Si substrate in potassium hydroxide (KOH) bath (6). Intermediate steps relate mainly to the challenge of preventing released nanobeams from collapsing due to their extreme aspect ratios. The 
	most important step, carried out prior to undercut step, involves recessing the Si substrate by $\sim$15 $\mu$m from the $\mathrm{Si_3N_4}$ layer via Bosch process(5). During the Bosch process the beams are protected using an upscaled version of the first e-beam mask (4). The final step is also crucial, in which critical point drying (CPD) is used to avoid structural collapse due to surface tension in of process drying the released structures.  Undercut takes place on individual $5\times12$ mm$^2$ sample chips diced from 700$\mu$m double sided polished Si wafer after step (3). The wafer is coated with a protective photoresist before dicing and to remove this protective layer and other organic 
	contaminants prior to undercut, sample chips are cleaned by NMP (1-methyl-2-pyrrolidon) and Piranha bath.

	\begin{figure*}[t!]
		\centering
		\includegraphics[width=2\columnwidth]{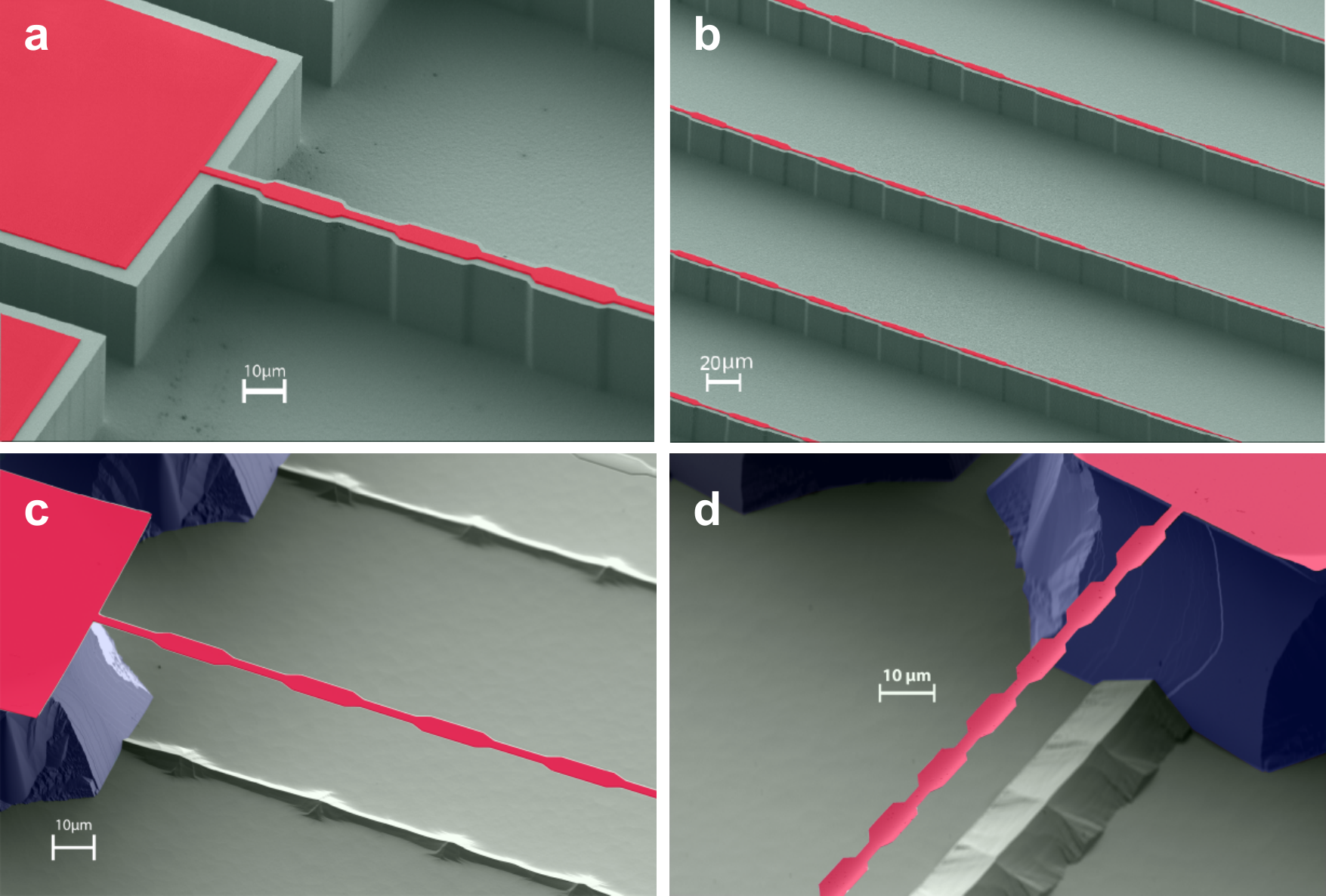}
		\caption{\textbf{False colored SEM images of the beams before and after releasing.} (a) and (b)  correspond to stage 5 of the process flow in Fig.~\ref{figure2}, focusing on (a) region near the supports and (b) tapered region of the beams. (c) and (d) correspond to the beams after KOH undercut. Red: $\mathrm{Si_3N_4}$, Blue: Silicon pillar, Green: Silicon substrate}
		\label{figureSEM}
		\vspace{-3mm}
	\end{figure*}
}

\end{document}